\documentclass[11pt,draftclsnofoot,onecolumn]{IEEEtran}
\usepackage{setspace}
\usepackage{clrscode}
\usepackage{amsthm}

\usepackage{amsmath}
\usepackage{amssymb}
\usepackage{graphicx}
\usepackage{epsfig}
\usepackage{hyperref}
\usepackage{pdfsync}
\usepackage{url}
\usepackage{color}

\usepackage{subfigure}

\newcommand{\tsnr}{\text{SNR}}

\newcommand{\tsinr}{\text{SINR}}

\newcommand{\figsize}{0.65}
\newcommand{\tmax}{\text{max}}

\newtheorem*{Lemnon}{Theorem}

\newtheorem{Rem}{Remark}

\begin{document}

\title{Broadbeam for Massive MIMO Systems}



%
\author{
{Deli Qiao, Haifeng Qian, and Geoffrey Ye Li}
\thanks{D. Qiao and Haifeng Qian are with the School of Information Science and Technology, East China Normal University, Shanghai, China, 200241 (e-mail: dlqiao@ce.ecnu.edu.cn, hfqian@cs.ecnu.edu.cn). Geoffrey Li is with the School of Electrical and Computer Engineering, Georgia Institute of Technology, Atlanta, Georgia 30332 (email:liye@ece.gatech.edu).}
\thanks{This work was supported in part by National Natural Science Foundation of China under Grant 61172085.}
}


\maketitle

\begin{abstract}
Massive MIMO has been identified as one of the promising disruptive air interface techniques to address the huge capacity requirement demanded by 5G wireless communications. For practical deployment of such systems, the control message need to be broadcast to all users reliably in the cell using broadbeam. A broadbeam is expected to have the same radiated power in all directions to cover users in any place in a cell. In this paper, we will show that there is no perfect broadbeam. Therefore, we develop a method for generating broadbeam that can allow tiny fluctuations in radiated power. Overall, this can serve as an ingredient for practical deployment of the massive MIMO systems.
\end{abstract}

\section{Introduction}

While the 4th generation (4G) wireless networks are vastly being deployed worldwide, 5G requirements and potential technologies have attracted the interest of both the academia and the industry recently. It is expected that 5G could address the massive capacity and massive connectivity challenges brought by the exponentially growing mobile traffic and machine type applications \cite{huawei-5G-whitepaper}. Massive MIMO systems \cite{tom-noncooperative} are equipped with a large number of transmit antennas at base stations and serve a large number of users simultaneous. It has been identified as a promising technique to address the challenges in 5G networks \cite{5gtechnique}.

In massive MIMO systems, the number of transmit antennas can be as large as hundreds or even thousands, which is a couple of orders larger than the current 4G systems (typically $4$ to $8$ antennas at most). The increase in the transmit antenna number can introduce many benefits, such as capacity, multiplexing, diversity, and energy efficiency. However, there are also many potential challenges for enabling massive MIMO \cite{lulu-mmimo}-\cite{resek-scalingmimo}. Precoding design is an important topic for realizing the benefits of massive MIMO sytems \cite{gcaire-massive} - \cite{lowcomp}. Well-designed precoding vectors can reduce the required antenna number or transmit power to achieve certain performance \cite{gcaire-massive}, and reduce the peak-to-average power ratio (PAPR) \cite{parlarsson}\cite{lowcomp}.

Generally, existing work on precoding design focuses on the current long-term evolution (LTE) systems \cite{tom-noncooperative}. However, there are still many problems open for practical deployment of massive MIMO compatible with LTE. One critical issue is how to design precoding to generate a reliable control channel, such as Physical Downlink Control Channel (PDCCH) and Physical Broadcast Channel (PBCH) \cite{ts36211}. Antenna virtualization has been used in \cite{samsung} to generate broadbeam. But the radiated power of the generated wide beam varies significantly in different directions.

On the other hand, beam pattern design has been an interesting topic for MIMO radar \cite{txbfmimoradar}-\cite{vorobyovtxbf}. For instance, in \cite{constmod}, an efficient optimization method for generating a constant modulus probing signal targeting a specific range of spacial angles has been proposed. In \cite{friedlander}, different methods for transmit beamforming have been investigated in MIMO radar based on the design of multiple correlated waveforms, where orthogonal waveforms and multi-rank transmit beamformer are combined to provide a general form of beamforming. Transmit beamspace techniques using multiple orthogonal waveforms for better direction finding have been proposed in \cite{vorobyovtxbf}. More recently, it has been shown in \cite{beamgen} that there are at most $2^{M-1}-1$ beamforming vectors generating the same beam pattern, where $M$ is the antenna number. Some of the methods for MIMO radar can be used to design precoding for control channels in cellular networks.

In this paper, we consider the specific problem of broadbeam generation for cellular systems with massive MIMO. We consider the uniform linear array (ULA) and the uniform rectangular array (URA) at the base station. Based on the idea in \cite{beamgen}, we first show that generation of perfect broadbeam can only result in trivial solutions. Therefore, there must be some fluctuations in the generated beam pattern. Then, we develop a method to find a precoding vector with negligible ripple and small peak-to-average power ratio (PAPR) or dynamic range (DR).

The organization of this paper is as follows. Section II presents the proposed method for generating broadbeam. In Section III, numerical examples are provided to show the effectiveness of the method. Finally, Section IV concludes this paper.

\section{Broadbeam Design}

\subsection{Uniform Linear Array}
\begin{figure}
\begin{center}
\includegraphics[width=\figsize\textwidth]{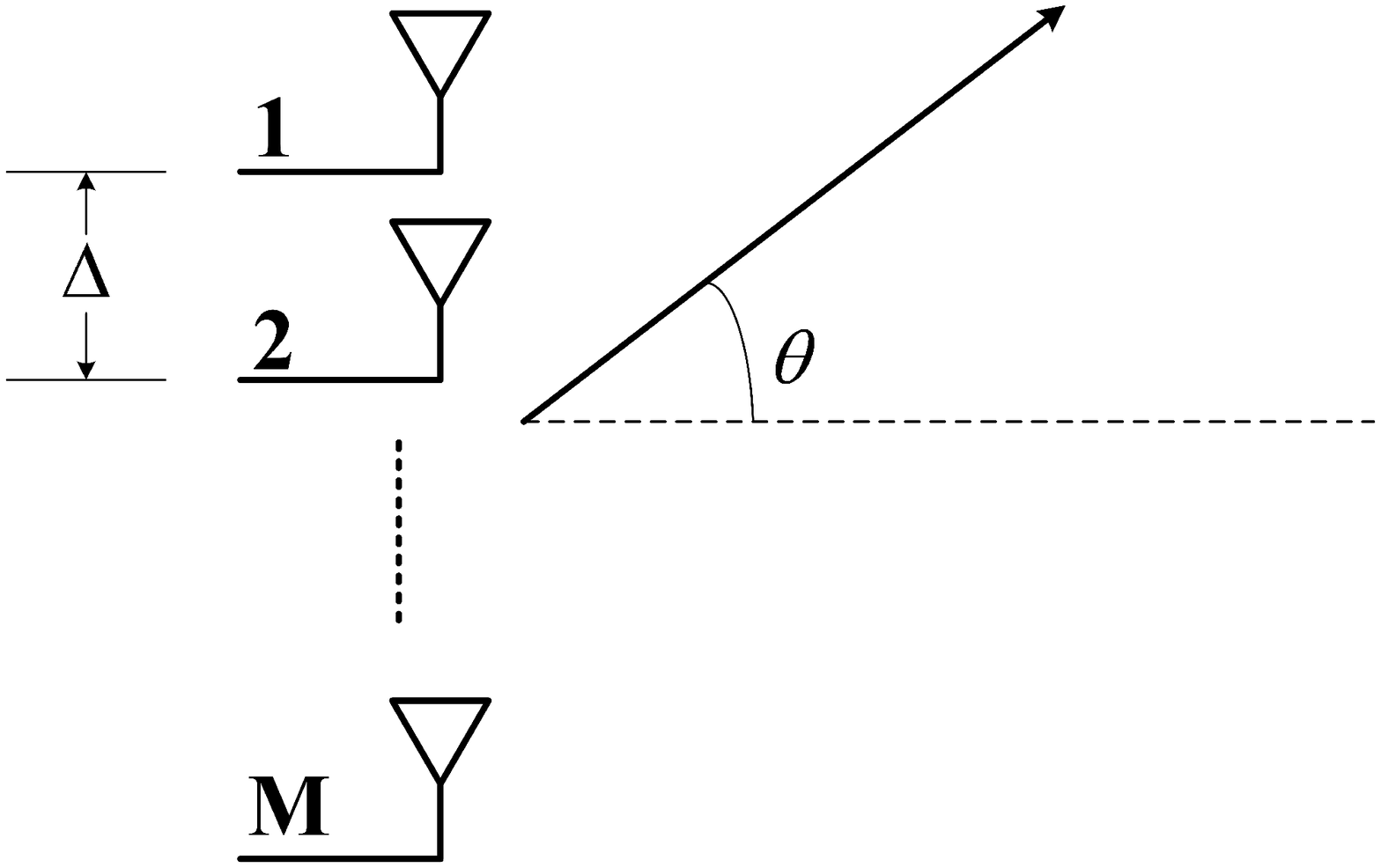}
\caption{ULA.}\label{fig:doa}
\end{center}
\end{figure}
Consider a base station with $M$ transmit antennas placed in a uniform linear array (ULA) as shown in Fig. \ref{fig:doa}, which is applicable in certain scenarios for massive MIMO systems \cite{mmimosurvey}. The steering vector towards angle $\theta$ has the form
\begin{align}\label{eq:steeringvec}
\mathbf{a}(\theta) = \left[1\quad a_2(\theta) \quad\cdots \quad a_M(\theta)\right]^T,
\end{align}
where $a_m(\theta)$, for $m=1,\ldots,M$, is a complex number representing the amplitude and phase shift of the signal at the $m$th antenna in relative to the first antenna with $a_1(\theta)=1$. $a_m(\theta)$ depends on the antenna structure. For ULA,
\begin{align}
a_m(\theta) = e^{j2\pi \frac{(m-1)\Delta}{\lambda} \sin(\theta)}
\end{align}
where $\Delta$ is the antenna spacing and $\lambda$ is the wavelength of the center frequency.

Let $\mathbf{v}=[v_1,\ldots,v_M]^T\in\mathbb{C}^{M\times1}$ denote the precoding vector for generating the broadbeam. The transmit beampattern generated is given by
\begin{align}\label{eq:beampatterndef}
f(\theta)= \mathbf{v}^H\mathbf{a}(\theta)\mathbf{a}^H(\theta)\mathbf{v}
\end{align}
where $[\cdot]^H$ denotes the matrix Hermitian. Then, the problem of designing a perfect broadbeam can be interpreted as
\begin{equation}\label{eq:problem1}
\begin{split}
\text{\textbf{P1}:}&\\
&\text{finding }\,\,\mathbf{v}\\
&\hspace{.6cm}\text{s.t. } f(\theta)=1, \forall \theta \in[-\frac{\pi}{2},\frac{\pi}{2}],\\
&\hspace{1.2cm}\text{and } \mathbf{v}^H\mathbf{v}=1.
\end{split}
\end{equation}

Based on the method indicated in \cite{beamgen}, we have the following result regarding the solutions to Problem \textbf{P1}.
\begin{Lemnon}
For an arbitrary ULA antenna of
size $M$, the only possible solutions for generating the perfect broadbeam are unit vectors, where only one element is of unit power while others are 0.
\end{Lemnon}

\emph{Proof: }Denote $\mathbf{D}(\theta)=\mathbf{a}(\theta)\mathbf{a}^H(\theta)$. Then,
\begin{align}\label{eq:beampattern-alt1}
f(\theta)=\mathbf{v}^H\mathbf{D}(\theta)\mathbf{v}.
\end{align}

Obviously, $\mathbf{D}(\theta)$ is a Toeplitz matrix. Denote the $2M-1$ elements generating the Toeplitz matrix $\mathbf{D}(\theta)$ as
\begin{align}\label{eq:diagonal-def}
\mathbf{w}(\theta)=\left[e^{-j2\pi(M-1)\frac{\Delta}{\lambda}\sin(\theta)},e^{-j2\pi(M-2)\frac{\Delta}{\lambda}\sin(\theta)},\ldots,1,e^{j2\pi\frac{\Delta}{\lambda}\sin(\theta)},\ldots,e^{j2\pi(M-1)\frac{\Delta}{\lambda}\sin(\theta)}\right]^T
\end{align}
 where $[\cdot]^T$ is the matrix transpose. In that case, the matrix, $\mathbf{D}(\theta)$, can be expressed as $\mathbf{T}(\mathbf{w}(\theta))$, with $\mathbf{T}(\cdot)$ the generator for Toeplitz matrix.

Now, consider the beampattern specified in Problem \textbf{P1} in (\ref{eq:problem1}). We can see that the radiated power is unit in all directions, that is, $f(\theta)=1$ for all $\theta$. To proceed, we need to choose a set of directions $\{\theta_1,\ldots,\theta_{2M-1}\}$ in $[-\frac{\pi}{2},\frac{\pi}{2}]$. Then, we can obtain the following set of equations
\begin{align}
\mathbf{v}^H\mathbf{T}(\mathbf{w}(\theta_k))\mathbf{v}=1,\forall k=1,2,\ldots,2M-1
\end{align}
The linear combinations of the above set of equations with arbitrary choice of coefficients $\mathbf{p}_i=[p_{i1},\ldots,p_{i,2M-1}]^T$ give us
\begin{align}\label{eq:weight-coeff}
\mathbf{v}^H\sum_{k=1}^{2M-1}p_{ik}\mathbf{T}(\mathbf{w}(\theta_k))\mathbf{v}=\sum_{k=1}^{2M-1}p_{ik}=\Sigma_i
\end{align}
where $\Sigma_i=\sum_{k=1}^{2M-1}p_{ik}$ is the sum of the elements of vector $\mathbf{p}_i$.

Since the Toeplitz matrix generator is a linear operation,
\begin{align}
\sum_{k=1}^{2M-1}p_{ik}\mathbf{T}(\mathbf{w}(\theta_k))=\mathbf{T}\left(\sum_{k=1}^{2M-1}p_{ik}\mathbf{w}(\theta_k)\right).
\end{align}
So, we can choose $\mathbf{p}_i$ such that all elements of the newly formed Toeplitz matrix are 0 except for the $i$-th diagonal elements. Note that, to make the diagonal elements of $\sum_{k=1}^{2M-1}p_{ik}\mathbf{T}(\mathbf{w}(\theta_k))$ become 0 except for the $i$-th diagonal, we only need to make sure that the element of $\sum_{k=1}^{2M-1}p_{ik}\mathbf{w}(\theta_k)$ are 0 except for the $i$-th one. Then, we can have the following set of equations
\begin{align}\label{eq:wprelation}
\left[\mathbf{w}(\theta_1)\,\mathbf{w}(\theta_2),\,\cdots,\, \mathbf{w}(\theta_{2M-1})\right]\mathbf{p}_i=\mathbf{e}_i,\,i=1,\ldots,2M-1.
\end{align}
where $\{\mathbf{e}_i\}$ are unit vectors with 1 for the $i$-th element and 0 else where.

Let $\mathbf{P}=\left[\mathbf{p}_1,\ldots,\mathbf{p}_{2M-1}\right]$, and $\mathbf{W}=\left[\mathbf{w}(\theta_1),\,\cdots,\, \mathbf{w}(\theta_{2M-1})\right]$, we now have
\begin{align}\label{eq:coeff-mtx}
\mathbf{WP}=\mathbf{I}
\end{align}
where $\mathbf{I}$ is the identity matrix. With the choice of $\{\theta_k\}$ given above, $\mathbf{W}$ is a Vandermonde matrix. So, as long as $\{\theta_k\}$ does not lead to overlapping elements in $\{e^{-j2\pi\frac{\Delta}{\lambda}\sin(\theta_k)}\}$, $\mathbf{W}$ will be invertible. Note that this can be achieved by carefully choosing $\{\theta_k\}$. To facilitate the proof, we assume $\theta=0$ is one sample. Without loss of generality, we choose $\{\theta_k\}$ such that $\sin(\theta_k)=\frac{2(k-M)}{2M-1},\,k=1,2,\ldots,2M-1$. Then, we have $\mathbf{P}=\mathbf{W}^{-1}$.

According to (\ref{eq:weight-coeff}) and (\ref{eq:wprelation}), we can have the following set of identities
\begin{align}
&v_1v_M^*=\Sigma_1,\label{eq:solutionset1}\\
&v_1v_{M-1}^*+v_2v_{M}^*=\Sigma_2,\\
&\hspace{2cm}\vdots\nonumber\\
&v_1v_2^*+v_2v_3^*+\cdots+v_{M-1}v_M^*=\Sigma_{M-1},\\
&|v_1|^2+\cdots+|v_M|^2=\Sigma_{M}\label{eq:solutionset2}
\end{align}
where $\Sigma_k$ is the sum of the elements of the $k$-th column of $\mathbf{W}^{-1}$. Denote the polynomial function
\begin{align}
\Xi_i(x)=\sum_{k=1}^{2M-1}p_{ik}x^k.
\end{align}
By decomposing (\ref{eq:coeff-mtx}), we have
\begin{align}
\Xi_i(e^{j2\pi\frac{\Delta}{\lambda}\left(\frac{2(k-M)}{2M-1}\right)})=\left\{\begin{array}{ll}0, &\forall k\neq i,\\ 1,& k=i.\end{array}\right.
\end{align}
Obviously, we can get
\begin{align}\label{eq:xires}
\Sigma_i=\Xi_i(1)=\left\{\begin{array}{ll}0,&\forall i\neq M,\\1,&i=M.\end{array}\right.
\end{align}
Substituting the above results to equations (\ref{eq:solutionset1})-(\ref{eq:solutionset2}), we can see that the only possible solutions are given by $|v_k|=1$ for some $k$ while $v_i=0, \forall i\neq k$, proving the theorem.\hfill$\square$

As shown in the above theorem, to achieve perfect broadbeam that radiates power identically in all directions, we can only let one antenna work. On the other hand, each antenna for massive MIMO systems should be inexpensive, lower power components \cite{mmimosurvey}. In this case, sending signal with only one antenna is extremely power inefficient, and fails to provide whole cell coverage.

If we allow the beampattern to fluctuate in the different directions within a very small amount, some useful broadbeams can be generated. Then, Problem \textbf{P1} can be modified into
\begin{equation}\label{eq:problem2}
\begin{split}
\textbf{P2}:&\\
&\text{finding } \,\,\mathbf{v}\\
&\hspace{.6cm}\text{s.t. }\, f(\theta)=1+\epsilon(\theta),\\
&\hspace{1.2cm} \text{and }\,\mathbf{v}^H\mathbf{v}=1.
\end{split}
\end{equation}
where $\epsilon(\theta)$ represents the fluctuation of the generated beampattern with bounded support, i.e., $|\epsilon(\theta)|\le \xi\ll 1$.

Denote $r_k=1+\epsilon(\theta_k)$ for the set of $\{\theta_k\}$ chosen in the above proof. Then, (\ref{eq:weight-coeff}) now becomes
\begin{align}
\mathbf{v}^H\sum_{k=1}^{2M-1}p_{ik}\mathbf{T}(\mathbf{w}(\theta_k))\mathbf{v}=\sum_{k=1}^{2M-1}r_k p_{ik}=\mathbf{r}^T\mathbf{p}_i
\end{align}
As a result, equations (\ref{eq:solutionset1})-(\ref{eq:solutionset2}) can be expressed as
\begin{align}
&v_1v_M^*=\mathbf{r}^T\mathbf{p}_1,\label{eq:bb-solutionset1}\\
&v_1v_{M-1}^*+v_2v_{M}^*=\mathbf{r}^T\mathbf{p}_2,\\
&\hspace{2cm}\vdots\nonumber\\
&v_1v_2^*+v_2v_3^*+\cdots+v_{M-1}v_M^*=\mathbf{r}^T\mathbf{p}_{M-1},\\
&|v_1|^2+\cdots+|v_M|^2=\mathbf{r}^T\mathbf{p}_{M.}\label{eq:bb-solutionset2}
\end{align}
Similar to \cite{beamgen}, the solutions to the above equation set can form
\begin{align}
g(x)&=\left(v_1+v_2x+\cdots+v_Mx^{M-1}\right)\left(v_1^*+v_2^*x^{-1}+\cdots+v_M^{*}x^{-(M-1)}\right),\label{eq:soluset1}\\
&=\mathbf{r}^T\mathbf{p}_1x^{-(M-1)}+\mathbf{r}^T\mathbf{p}_2x^{-(M-2)}+\cdots+\mathbf{r}^T\mathbf{p}_M+\mathbf{r}^T\mathbf{p}^*_{M-1}x+\cdots+\mathbf{r}^T\mathbf{p}_1^*x^{M-1}.\label{eq:soluset2}
\end{align}

From the structure of $g(x)$ in (\ref{eq:soluset1}), if $x_1,x_2,\ldots,x_{M-1}$ are solutions to $g(x)=0$, then $\frac{1}{x_1^*},\frac{1}{x_2^*},\ldots,\frac{1}{x_{M-1}^*}$ are too. From the solution set of $g(x)=0$, we can form at most $2^{M-1}$ solutions of Problem \textbf{P2} by
\begin{align}\label{eq:precodingvec}
\phi(x)=\prod_{m=1}^{M-1}(x-\alpha_m)=v_1+v_2x+\cdots+v_Mx^{M-1},
\end{align}
where $\alpha_m=x_m$ or $\frac{1}{x_m^*}$.

Even though Problem \textbf{P2} has at most $2^{M-1}$ solutions, we are interested in the one that achieves the lowest possible PAPR defined as
\begin{align}\label{eq:papr}
\delta=\frac{M\max_{m}|v_m|^2}{\parallel\mathbf{v}\parallel^2}
\end{align}
for the precoding vector $\mathbf{v}$. Since there are $2^{M-1}$ possible precoding vectors, exhaustive search can be used to find it. 
%

\begin{Rem}
With the above characterization, we can see from (\ref{eq:xires}) that $g(x)\equiv 1$ for perfect broadbeam generation, i.e., there is no solution to $g(x)=0$. Therefore, we can not find $2^{M-1}$ solutions to Problem \textbf{P1} as we do for Problem \textbf{P2}.
\end{Rem}

To summarize the above discussions, we propose the following procedures on the top of the next page to obtain the desired precoding vector for generating broadbeam. Note that given the antenna setting, the previous algorithm can be performed offline. Hence, complexity is not a problem.

\begin{figure*}
\begin{codebox}
\Procname{$\proc{Broadbeam Generation Method (BGM)}$}
 \li      Choose $\epsilon(\theta)$;
\li       Choose $\theta_k,\,k=1,\ldots,2(M-1)$ such that $\sin(\theta_k)=\frac{2(k-M)}{2M-1}$;
 \li      Obtain the matrix $\mathbf{W}=\left[\mathbf{w}_1,\ldots,\mathbf{w}_{2M-1}\right]$, and the corresponding beam pattern vector \li $\mathbf{r}=[1+\epsilon(\theta_1),\ldots,1+\epsilon(\theta_{2M-1})]^T$;
    \li   Solve the polynomials defined in (\ref{eq:soluset2}) for solutions $\{x_1,\ldots,x_{M-1},1/x_1^*,\ldots,1/x_{M-1}^*\}$;
\li     Loop over all possible pairs of solutions to find the associated precoding vector $\mathbf{v}$ specified in (\ref{eq:precodingvec});
\li    In the loop, save the precoding vector $\mathbf{v}$ with lowest PAPR defined in (\ref{eq:papr});
 \li   The desired precoding vector is given by $\frac{\mathbf{v}}{\parallel\mathbf{v}\parallel}$.
\end{codebox}
\hrule
\end{figure*}

\subsubsection{Peak Power Constraint}
Moreover, in practical use, the antennas may be subject to a peak power constraint, i.e., $|v_m|^2\le v_{\tmax},\,m=1,2,\ldots,M$. Then, we need to normalize the precoding vector as follows
\begin{align}
\frac{\mathbf{v}}{\max_m |v_m|}\sqrt{v_{\tmax}}.
\end{align}
In this case, the base station should radiate the power as much as possible. That is, we need to find a precoding vector with maximum radiated power, i.e., $\parallel\mathbf{v}\parallel^2$. Note that this problem is equivalent to finding a precoding vector with minimum PAPR defined in (\ref{eq:papr}).

\subsubsection{Dynamic Range}
Another metric of interest is the dynamic range (DR), which is defined as
\begin{align}
\frac{\max_m|v_m|^2}{\min_m|v_m|^2}.
\end{align}
In this case, we would like to minimize the dynamic range. Compared with (\ref{eq:papr}), we can see that the difference with PAPR-based method lies in the denominator, which is now the minimum power of the antennas.

\subsection{Uniform Rectangular Array}
\begin{figure}
\begin{center}
\includegraphics[width=\figsize\textwidth]{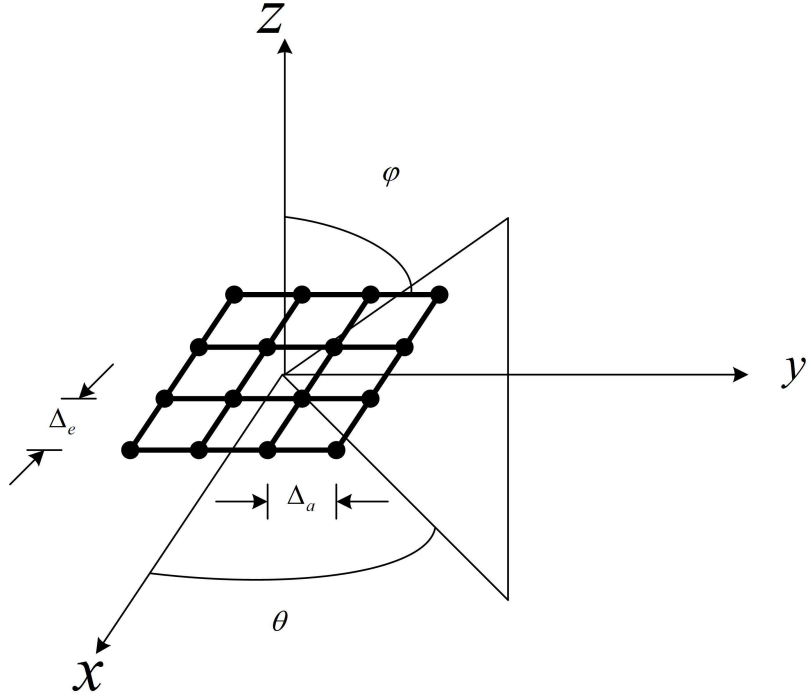}
\caption{Uniform rectangular array.}\label{fig:UPA}
\end{center}
\end{figure}
Note that generating broadbeam for URA is similar to the case of ULA except that we need to consider the azimuth and elevation angles. Consider a uniform rectangular array with $M\times N$ identical antennas placed with uniform spacing as shown in Fig. \ref{fig:UPA}. The component of the steering vector for each antenna in the direction $(\psi,\theta)$ is given by \cite{antennabook}
\begin{small}
\begin{align}
\hspace{-.6cm}[\mathbf{A}(\psi,\theta)]_{mn}=a_{mn}(\psi,\theta) = e^{-j2\pi \frac{(m-1)\Delta_a}{\lambda} \sin(\psi)\sin(\theta) -j2\pi \frac{(n-1)\Delta_e}{\lambda} \sin(\psi)\cos(\theta)}.
\end{align}
\end{small}
Then, the steering vector can be written as
\begin{align}
\mathbf{a}(\psi,\theta) = \text{vec}(\mathbf{A}(\psi,\theta)) = \mathbf{a}_a(\psi,\theta)\otimes \mathbf{a}_e(\psi,\theta)
\end{align}
where $$
\mathbf{a}_a(\psi,\theta) = [1,e^{j2\pi \frac{\Delta_a}{\lambda} \sin(\psi)\sin(\theta)},\ldots,e^{j2\pi \frac{(M-1)\Delta_a}{\lambda} \sin(\psi)\sin(\theta)}]^T$$
 and $$\mathbf{a}_e(\psi,\theta) = [1,e^{j2\pi \frac{\Delta_e}{\lambda} \sin(\psi)\cos(\theta)},\ldots,e^{j2\pi \frac{(N-1)\Delta_e}{\lambda} \sin(\psi)\cos(\theta)}]^T.$$

If we let $\mathbf{v}=\mathbf{v}_a\otimes\mathbf{v}_e$, the transmit beam pattern can be expressed as
\begin{small}
\begin{align}
f(\psi,\theta) &= \mathbf{v}^H\mathbf{a}(\psi,\theta)\mathbf{a}^H(\psi,\theta)\mathbf{v}\\
& = (\mathbf{v}_a\otimes\mathbf{v}_e)^H(\mathbf{a}_a(\psi,\theta)\otimes \mathbf{a}_e(\psi,\theta))\nonumber\\
&\hspace{.5cm}\cdot(\mathbf{a}_a(\psi,\theta)\otimes \mathbf{a}_e(\psi,\theta))^H(\mathbf{v}_a\otimes\mathbf{v}_e)\\
& = (\mathbf{v}^H_a\otimes\mathbf{v}^H_e)(\mathbf{a}_a(\psi,\theta)\otimes \mathbf{a}_e(\psi,\theta))\nonumber\\
&\hspace{.5cm}\cdot(\mathbf{a}^H_a(\psi,\theta)\otimes \mathbf{a}^H_e(\psi,\theta))(\mathbf{v}_a\otimes\mathbf{v}_e)\\
& = (\mathbf{v}^H_a\mathbf{a}_a(\psi,\theta)\mathbf{a}^H_a(\psi,\theta)\mathbf{v}_a)\nonumber\\
&\hspace{.5cm}\otimes(\mathbf{v}^H_e\mathbf{a}_e(\psi,\theta)\mathbf{a}^H_e(\psi,\theta)\mathbf{v}_e)\\
& = (\mathbf{v}^H_a\mathbf{a}_a(\psi,\theta)\mathbf{a}^H_a(\psi,\theta)\mathbf{v}_a)\nonumber\\
&\hspace{.5cm}\cdot(\mathbf{v}^H_e\mathbf{a}_e(\psi,\theta)\mathbf{a}^H_e(\psi,\theta)\mathbf{v}_e)\label{eq:UPAdec-proof1}\\
& = f_a(\psi,\theta)f_e(\psi,\theta)\label{eq:UPAdec}
\end{align}
\end{small}
where the following properties of Kronecker product are used: 1) $(A\otimes B)^H=A^H\otimes B^H$; 2) $(A\otimes B)(C\otimes D)=(AC)\otimes (BD)$, and (\ref{eq:UPAdec-proof1}) holds since both terms inside the parenthesis are scalar values, $f_a(\psi,\theta) = \mathbf{v}^H_a\mathbf{a}_a(\psi,\theta)\mathbf{a}^H_a(\psi,\theta)\mathbf{v}_a$, and $f_e(\psi,\theta) =\mathbf{v}^H_e\mathbf{a}_e(\psi,\theta)\mathbf{a}^H_e(\psi,\theta)\mathbf{v}_e$.

Then, similar to (\ref{eq:problem1}), we can interpret the problem of designing broadbeam for URA as
\begin{equation}\label{eq:problem3}
\begin{split}
\text{\textbf{P3}:}&\\
&\text{finding }\,\,\mathbf{v}\\
&\hspace{.6cm}\text{s.t. } f(\psi,\theta)=1, \forall \psi\in[-\frac{\pi}{2},\frac{\pi}{2}],\theta \in[-\frac{\pi}{2},\frac{\pi}{2}],\\
&\hspace{1.2cm}\text{and } \mathbf{v}^H\mathbf{v}=1.
\end{split}
\end{equation}
Combining (\ref{eq:UPAdec}), we can decompose the previous problem into two subproblems of finding $\mathbf{v}_a$ and $\mathbf{v}_e$ with $f_a(\psi,\theta)=1$ and $f_e(\psi,\theta)=1$ as constraints, respectively. Note that they are similar to the discussions for ULA, and hence the \textbf{Theorem} holds for URA as well. Also we can design precoding vectors that can allow some fluctuations in radiation pattern following similar steps.

\section{Numerical Results}

\subsection{Beam Pattern}

In this part, we will evaluate the proposed method for generating broadbeam. Note that the algorithm proposed is generic. Since $2^{M-1}$ is very large for a large number of antennas, due to the limitation in computing resource, we here only show the results for the case $M=16$ with $\Delta=\frac{1}{2}\lambda$ for ULA, while we assume $8\times8$ array for URA. The precoding vector for larger number of antennas can be derived similarly.
We consider two different performance metrics, PAPR and dynamic range (DR). In the following figures, ``PAPR-based'' refers to the method with PAPR as the optimization metric, while ``DR-based'' refers to the method with dynamic range as an optimization metric.

\begin{figure}
\centering
\includegraphics[width=\figsize\textwidth]{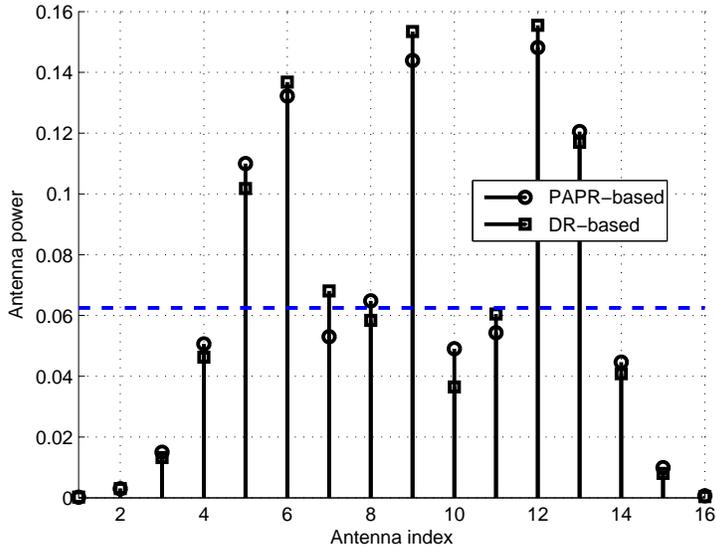}
\caption{Power of each antenna. $M=16$. $\xi=0.01$.}\label{fig:M=16xi=001coeff}
\end{figure}
\begin{figure}
\begin{center}
\includegraphics[width=\figsize\textwidth]{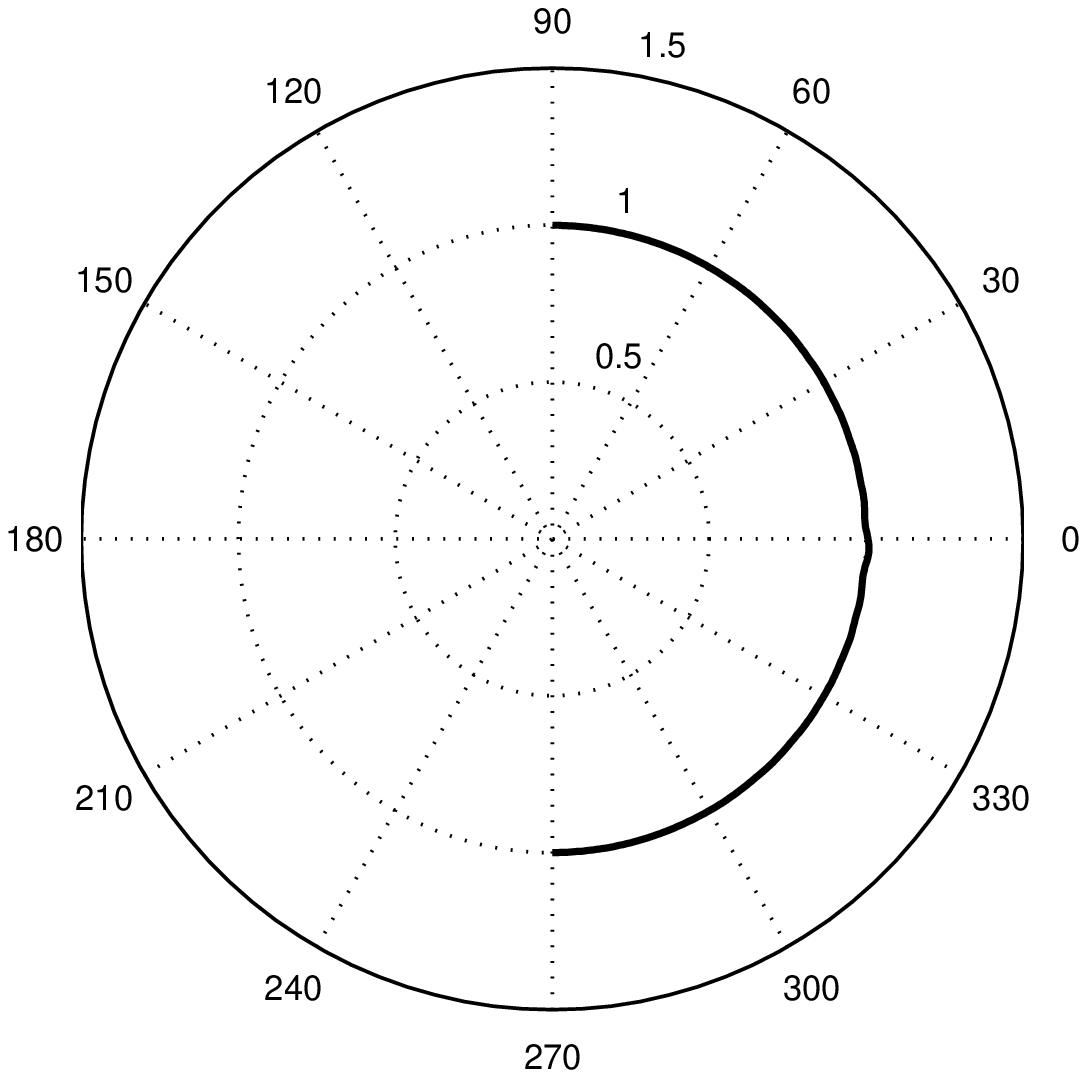}
\caption{Beam pattern. PAPR-based. $M=16$.}\label{fig:M=16xi=001pattern}
\end{center}
\end{figure}

Consider a ULA antenna array with 16 antennas. Assume $\xi=0.01$.
The PAPR obtained is $\delta=2.37=3.75$ dB while the minimum dynamic range is 28 dB. In Fig. \ref{fig:M=16xi=001coeff}, we plot the corresponding power of each antenna. In the figure, the circles and squares represent the power of different antennas with PAPR-based or DR-based optimization method, where the dashed line represent the perfect scenario with PAPR $\delta=1=0$ dB, i.e., constant envelope with $\frac{1}{16}$. In Fig. \ref{fig:M=16xi=001pattern}, we plot the associated beam pattern for PAPR-based method. We do not provide the beam pattern associated with DR-based method, since it is hardly distinguishable from the one with PAPR-based method. As can be seen from the figure, the beam pattern for the generated precoding vector is almost flat for $[-90^o,90^o]$.

\begin{figure}
\begin{center}
\includegraphics[width=\figsize\textwidth]{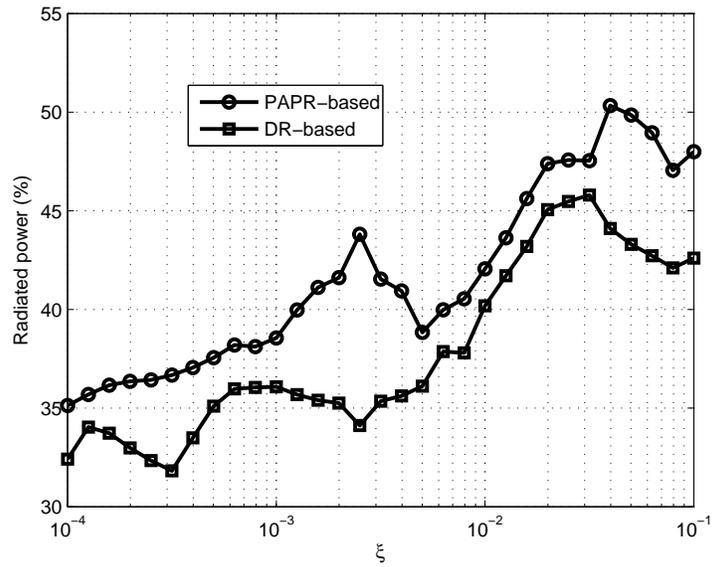}
\caption{Total radiated power v.s. $\xi$. $M=16$.}\label{fig:paprinxi}
\end{center}
\end{figure}

\begin{figure}
\begin{center}
\includegraphics[width=\figsize\textwidth]{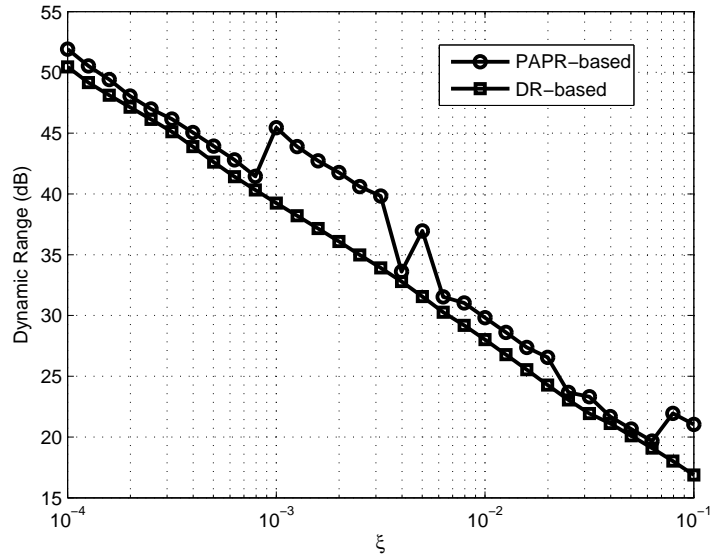}
\caption{Dynamic range v.s. $\xi$. $M=16$.}\label{fig:drinxi}
\end{center}
\end{figure}

\begin{figure}
\begin{center}
\includegraphics[width=\figsize\textwidth]{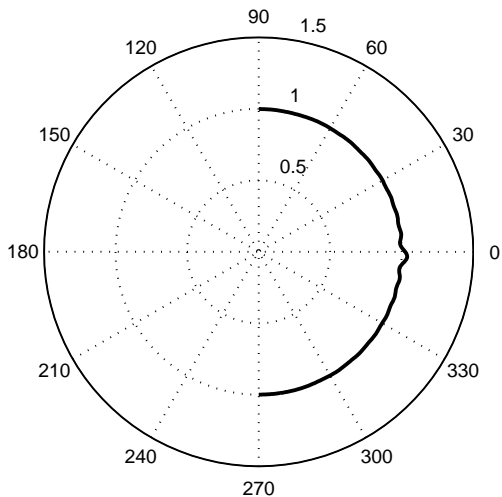}
\caption{Beam pattern. PAPR-based. $M=16$.}\label{fig:M=16xi=004pattern}
\end{center}
\end{figure}

We are also interested in the total radiated power with the broadbeam in the presence of additional peak power constraints on antennas. Here, we assume that the peak power is $v_{\tmax}=\frac{1}{M}=\frac{1}{16}$. We plot the total radiated power in percentage with respect to the full power of 1 as a function of $\xi$ in Fig. \ref{fig:paprinxi}. From the figure, the overall trend for radiated power is increasing in $\xi$, since smaller $\xi$ requires the antennas to counteract the interactions between each other more stringently, which generally wastes more power. Note that if we send signal with only single antenna, the power radiated is $\frac{1}{16}=6.25\%$. We can obtain a significant boost in radiated power with the proposed method, e.g., 9 dB increase at $\xi=0.04$ where around $50\%$ of the total power can be radiated, and hence in the coverage range. And it is not surprising that reducing dynamic range wastes more power. In addition, we plot the dynamic range as a function of $\xi$ in Fig. \ref{fig:drinxi}. From the figure, we can see that the dynamic range is decreasing in $\xi$. It is interesting that for every 10 dB decrease in the ripple of generated broadbeam, the increment in dynamic range is by around 10 dB. This provides us a tradeoff between fluctuations in radiated pattern and dynamic range of antennas. Moreover, it is interesting that there are some local maximum of radiated power with respect to $\xi$. For instance, there is a local maximum around $\xi=0.04$ with more than half of the total power radiated for PAPR-based method. So, we also plot the associated beam pattern for PAPR-based method in Fig. \ref{fig:M=16xi=004pattern}. Compared with Fig. \ref{fig:M=16xi=001pattern}, we can find that the increase in radiation power is at the expense of larger spikes in beam pattern, which may introduce larger inter-cell interference in cellular systems.

\begin{figure*}
\centering
  \subfigure[Beam pattern.]{
    \label{fig:UPA-a} 
    \includegraphics[width=0.5\textwidth]{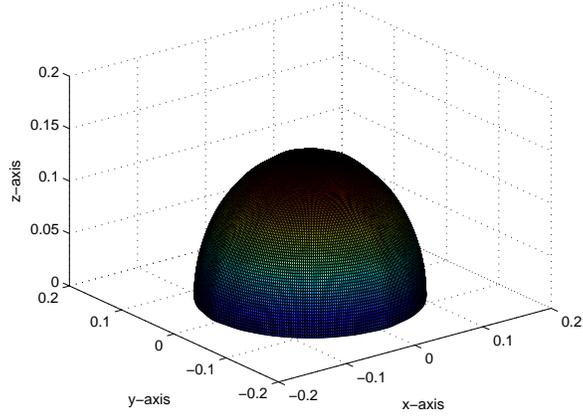}}
  \subfigure[Azimuth pattern.]{
    \label{fig:UPA-b} 
    \includegraphics[width=0.5\textwidth]{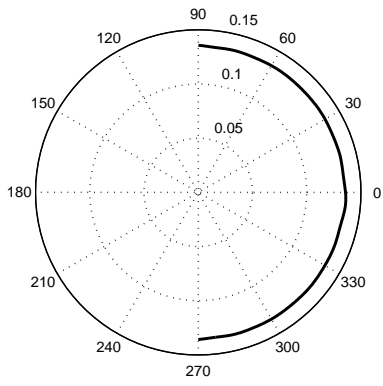}}
 \subfigure[Elevation pattern.]{
    \label{fig:UPA-c} 
    \includegraphics[width=0.5\textwidth]{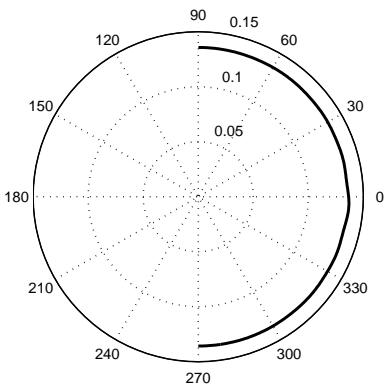}}
  \caption{Beam pattern for $8\times8$ URA.}

\end{figure*}

So far, we have provided results for ULA. In Fig. \ref{fig:UPA-a}, we plot the beam pattern for a $8\times8$ uniform rectangular array with $\Delta_a=\Delta_e=\frac{\lambda}{2}$, with Fig. \ref{fig:UPA-b} and \ref{fig:UPA-c} representing azimuth and elevation pattern for illustration. We can see from the figures that the proposed method can apply to the uniform rectangular array as well.

\subsection{CDF of SINR}
Similar to \cite{tom-noncooperative}, we consider a typical cellular network with 19 cells. We assume ULA with half wavelength spacing for each base station located at the center of each cell. Assume that the cell radius is 1.6km, the antenna number is $M=16$, the number of users per cell is 10, and no user is located within 100m of the base stations. The parameters for simulation are summarized in Table \ref{tab:parameters}. Regarding the channel model, we assume that the channel formed between user $k$ in cell $l$ and the BS in cell $l'$ is given by \cite{tom-noncooperative}
\begin{align}
\mathbf{h}_{kll'}=\sqrt{\beta_{kll'}}\mathbf{g}
\end{align}
where $\mathbf{g}\in\mathbb{C}^{M\times1}\sim\mathcal{CN}(\mathbf{0},\mathbf{I})$ denotes the fast Rayleigh fading coefficients, and $\beta_{kll'}$ denotes the path loss with
\begin{align}
\beta_{kll'}=\frac{1}{d_{kll'}^\gamma}
\end{align}
where $d_{kll'}$ is the distance between the base station $l'$ and user $k$ in cell $l$, and $\gamma\in(2,4)$ is the path loss exponent. Note that the log-normal fading is not considered here.

\begin{table}
\begin{center}
\begin{tabular}{ |l|l| }
  \hline
  BS power ($P$) & 46 dbm\\ \hline
  System Bandwidth ($B$) & 20MHz \\ \hline
  Noise power density ($N_0$) & -174dbm/Hz\\ \hline
  Cell Radius  & 1600 m \\ \hline
  Cell Hole & 100m\\ \hline
  BS Antenna number  & 16 \\ \hline
  Antenna Configuration & $ULA$ \\ \hline
  Antenna Separation & $\frac{\lambda}{2}$ \\ \hline
  Number of UEs per Cell & 10\\ \hline
  UE Antenna number & 1 \\
  \hline
\end{tabular}
\end{center}
\caption{Simulation parameters.}\label{tab:parameters}
\end{table}

We are interested in the received signal-to-interference-and-noise ratio (SINR). For the baseline, we consider the \tsinr\,obtained as if the base station is only equipped with one antenna sending signals with full power, termed as ``Geometry''. In this case, the received \tsinr\, of user $k$ in cell $l$ is defined as
\begin{align}
\tsinr_{kl,geo}=\frac{\tsnr\beta_{kll}|h_{kll}|^2}{1+\tsnr\sum_{l'\neq l}\beta_{kll'}|h_{kll'}|^2}
\end{align}
where $\tsnr=\frac{P}{N_0B}$ denotes the transmit signal-to-noise ratio (SNR).
If we assume that the base stations send signals with broadbeam in each cell, the received signal of user $k$ in cell $l$ for each symbol can be expressed as
\begin{align}
y_{kl} =  \underbrace{\mathbf{h}_{kll}^H\mathbf{v}s_{l}}_{\text{useful signal}} + \underbrace{\sum_{l'\neq l} \mathbf{h}_{kll'}^H\mathbf{v}s_{l'}}_{\text{inter-cell interference}}+ n_{kl}
\end{align}
where $\mathbf{v}$ denotes the precoding vector generating the broadbeam, $s_l\sim\mathcal{CN}(0,P/B)$ denotes the signal sent by the base station in cell $l$, $n_{kl}\sim\mathcal{CN}(0,N_0)$ is the circularly symmetric complex Gaussian noise at the user side.
Now, we can obtain the received signal-to-interference-and-noise ratio (\tsinr) of user $k$ in cell $l$ as
\begin{align}
\tsinr_{kl}=\frac{\tsnr|\mathbf{h}^H_{kll}\mathbf{v}|^2}{1+\tsnr\sum_{l'\neq l}|\mathbf{h}^H_{kll'}\mathbf{v}|^2}.
\end{align}

\begin{figure}
\begin{center}
\includegraphics[width=\figsize\textwidth]{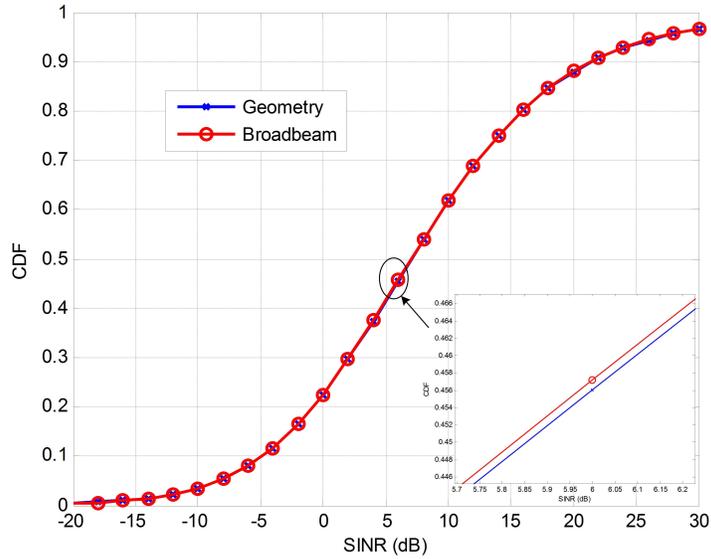}
\caption{CDF of received SINR.}\label{fig:sinrcdf}
\end{center}
\end{figure}

In simulations, we assume that the UEs are uniformly distributed over the coverage area. We consider 10 drops, each with $10^3$ instances of the channel coefficients. In Fig. \ref{fig:sinrcdf}, we plot the cumulative distribution function (CDF) of \tsinr\, with parameters defined in Table \ref{tab:parameters}. As can be seen from the figure, our proposed broadbeam generation method can achieve performance close to the one as if only one antenna sending with full power.

\section{Conclusions}

In this paper, we have considered broadbeam generation in massive MIMO systems. We have shown that the only possible solutions to perfect broadbeam with identical radiated power in all directions are the unit vectors with only one nonzero element. By allowing some fluctuations, we have proposed a method to generate broadbeam that is almost flat in all directions while minimizing the PAPR or dynamic range for practical applications. We have also provided numerical results verifying our algorithm. Overall, we have offered a feasible solution to generating broadbeam of practical use in massive MIMO systems.


\end{document}